*Review*

# Synthetic Media in Multilingual MOOCs: Deepfake Tutors, Pedagogical Effects, and Ethical - Policy Challenges

**Alexandros Gazis[1,*], Erietta Chamalidou[2], Nikolaos Ntaoulas[3] , Theodoros Vavouras[4,5]**

[1] *Department of Electrical and Computer Engineering, School of Engineering, Democritus University of Thrace, 67100 Xanthi, Greece;* agazis@ee.duth.gr

[2] *Department of Electrical and Computer Engineering, School of Engineering, National and Technical University of Athens, 15773, Greece;* el24613@ntua.gr

[3] *Ministry of Education, Religious Affairs and Sports, 15122 Athens, Greece;* nida@sch.gr

[4]*Department of Humanities, School of Humanities, Hellenic Open University, 26335 Patra, Greece;* vavouras.theodoros@ac.eap.gr

[5] *Department of Philosophy, School of Italian Language and Literature, Aristotle University of Thessaloniki, 54124 Thessaloniki, Greece;* vavouras@itl.auth.gr

* Correspondence: agazis@ee.duth.gr

**Abstract:** In recent years, synthetic media from deepfake videos have emerged as a new interesting technology, whether that refers to cloned voices, multilingual translation models, or more recent applications of avatar tutors into higher education. As such, these technologies are rapidly becoming part of the multilingual distance learning model and, more recently, MOOCs worldwide. On the one hand, synthetic media is a complex term often used to describe risks, where, on the other hand, recent studies indicate that when used transparently and with a focus on pedagogical and not entertainment objectives, they can improve accessibility, personalization, and increase learners' engagement. This article is a scoping review that focuses on recent international literature published between 2020 and 2025 to explore the usage of deepfake and synthetic media tools and methods in multilingual MOOC content and assess the influence of these technologies on social presence and participation. Similarly, we focus on ethical and political issues that are closely connected with the adaptation of these technologies, and upon analysing educational technology and policy documents, such as UNESCO's Guidelines and the EU AI Act, we pinpoint that the use of synthetic avatars and AI-generated videos can diminish production costs and assist multilingual learning. Evidently, concerns arise regarding authenticity, privacy, and the shifting nature of the teacher-learner relationship that are thoroughly discussed. As a result, the technical merit of this paper is the proposal of a policy framework that, in an effort to address these issues, focuses on transparency, responsible governance, and AI literacy. The goal is not to replace human instruction but to integrate synthetic media in ways that strengthen pedagogical design, safeguard rights, and ensure that multilingual MOOCs become more interesting

---

[1] Article Affiliation
Author email address

and inclusive rather than more automated robotic processes and unequal.

# 1. Introduction

The rapid development of generative artificial intelligence models between 2020 and 2025, including large language models (LLMs), multimodal translation systems, and voice and video synthesis technologies, has created an entirely new synthetic media ecosystem, [1]. These media now extend far beyond basic image filters and include deepfake videos, neural translation systems, avatar tutors, and AI-generated educational videos that can be adapted to different languages, tones, speaking rates, and even cultural contexts. In higher education, and particularly in Massive Open Online Courses (MOOCs), such technologies are especially attractive because they enable more cost-effective production of multilingual material, greater flexibility in creating and updating educational videos, personalised learning through avatars or synthetic "teachers," and improved access for students who do not speak English as a first language, [2,3]. Yet the broader use of deepfakes and synthetic media also raises concerns about the authenticity of educational content, the transparency of whether an instructor is human or synthetic, the protection of personal data such as voice and image, the preservation of academic integrity, and the potential for deception, [4-6].

To date, most discussions on deepfakes in higher education have focused on risks such as misinformation, reputation damage, and privacy violations, [7-11]. Specifically, in study, [12], the authors note that areas such as politics, security, or social media are not examined in detail regarding the educational application domain. Similarly, newer studies such as [13], focus more on AI tutors in higher education, thus showcasing that students can have similar or even improved learning outcomes in comparison with "old-fashioned" traditional video instructors. In this article, it is noted that the sense of trust and the final perceived relation with the "digital instructor" are not the same as expected. Likewise, in [14], the authors presented a study of 447 participants that suggests learning from AI-generated educational videos can be compared with video learning featuring human tutors, but students were noticed to experience lower social presence and fewer positive emotional responses and connection when exposed to full-fledged synthetic videos. The results of these studies are that Generative AI is evidently influencing both the technical phase of online education and the pedagogical one alike, meaning it affects pedagogical notions such as presence, authenticity, and multilingualism.

In international literature, the term deepfake refers to generatively produced or modified audiovisual content that imitates a real person's appearance or voice with high realism, using deep learning techniques such as GANs and diffusion models, [15-17]. The broader category of synthetic media includes synthetic speech, [18], synthetic video, [19], synthetic images, [20], synthetic text, [21], and multimodal formats such as speech-to-speech translation, [22], that generate new audio in another language, [23], [24]. In this paper, the terms deepfake and synthetic media refer not to malicious applications but to educational uses, similar to [13,25], that build on the same technological principles, are deployed in distance-learning contexts such as MOOCs, [26], and propose forms of pedagogical innovation, [27] such as multilingual avatars, [28] or adaptive narration, [29,30]. This distinction is important because public debate commonly associates deepfakes with manipulation, whereas this study focuses on their educational potential and the ethical requirements that accompany responsible implementation.

MOOCs were originally intended to democratise access to knowledge, [32], yet the predominance of English has produced linguistic and cultural inequalities, [33,34]. A lot of non-English speaking individuals from diverse backgrounds find it hard to attend or follow in general advanced courses, even though the model's declared openness. In contrast, newer technology innovations such as SEAMLESSM4T, a framework that allows simultaneous speech and text translation in up to 100 languages, [35-38], prove that it is technically feasible to provide live translations with captions (subtitles) to educational videos automatically, and if needed on a large scale. As such, the integration of these systems in MOOC systems provides a new opportunity for one instructor to teach in multiple languages while at the same time maintaining their unique characteristics and, more evidently, their voice, facial expression, and overall presence. Also, alternative videos can be offered in the original language via caption, i.e., adapted subtitles or generic simplified language. Additionally, recent studies on AI-generated videos in the domain of education [39-43], suggest that the overall learning outcomes are comparable to real-life human-produced material, while at the same time decisive factors remain notions such as social presence and authenticity. For this reason, multilingual MOOCs enhanced with synthetic media have the potential to support greater linguistic equity, provided their use is guided by clear pedagogical, ethical, and political principles, [44,45].

A full understanding of synthetic media in education requires engagement with pedagogical and ethical frameworks rather than purely technical descriptions. The Community of Inquiry (CoI) model is relevant here, [46], as synthetic avatars and AI-generated videos influence the balance of cognitive, social, and teaching presence, [47,48]. Social presence may be altered when the "instructor" is an avatar; teaching presence shifts as instructors increasingly take on roles of designers or curators; and cognitive presence changes as the representation, pacing, or linguistic form of information is modified. Heutagogical perspectives also offer insight, emphasising autonomous and self-directed learning in which students can select the language, avatar, or difficulty level of learning resources, participate in the co-creation of instructional material, and develop critical awareness of AI and deepfake technologies, [49-51]. Ethical considerations are also important to study and note. For example, the authors in [52], argue regarding the ethics of educational AI and whether it should be treated as a transparent process involving all interested parties participating in evaluating its use and adaptation. Similarly, [53] focuses on notions like technological maturity, transparency, reproducibility, and privacy of LLMs, which become increasingly complex when fused with multimodal synthetic media, [54]. As such, one can state that MOOCs and synthetic media should be studied as standalone pedagogical and ethical terms rather than mere tools, methods, or technical solutions.

Another interesting issue is the debate that exists in correlation with international and European regulatory frameworks. Specifically, UNESCO's Guidance for Generative AI in Education and Research, published in 2023, [55], clearly states principles for responsible adoption, emphasising human-centred design, safety, content transparency, digital and AI literacy, and the safeguarding of privacy and human rights. On the other hand, Europe has also regulated the AI Act (Regulation (EU) 2024/1689), [56], which presents risk categories for AI systems and defines the required assessment criteria for high-risk applications, one of which is the educational domain, [57]. Finally, it is worth noting that although deepfakes in education are not explicitly addressed per se, its provisions on transparency, risk management, and consistent market monitoring are directly relevant to synthetic media in MOOCs. As a result, these regulations provide an overall framework for operation for educational institutions to stand and develop their own internal policies on synthetic media and legal procedures for informed consent. This is particularly interesting and important as, for example, when real educators' voices and/or images are used, one should provide full disclosure of all possible synthetic content and use, thus promoting accountability and public dialogue regarding the implementation of generative AI in education.

## 2. Aims and Objectives

This paper aims to present a literature review on the different ways deepfakes and synthetic media are used for multilingual MOOC content. As such, our analysis focuses on the pedagogical and ethical implications.

Three core *research questions* guide the study:

1. **RQ1.** What are the deepfake and synthetic media forms? For example, MOOC solutions may consist of AI tutors, avatars, voice cloning, text-to-video systems, or speech-to-speech translation.
2. **RQ2.** What are the pedagogical implications of the above-mentioned technologies? For instance, one must study and understand the impact on the students' learning performance, engagement, social presence, and perceptions of authenticity.
3. **RQ3.** What are the key ethical, cultural, and political challenges? For example, terms like transparency, privacy, misinformation, and access inequalities should be taken into consideration in regard to the proposed policy directions of MOOCs' technological advancements.

The *contribution* of this work is fourfold:

1. **C1.** It fuses pedagogical and ethical dimensions, using synthetic media as a site for embedding ethics and politics rather than as mere technical tools.
2. **C2.** It foregrounds multilingual MOOCs rather than generic AI trends in education, promoting access, cultural adaptation, and linguistic justice, and providing concrete examples such as SEAMLESSM4T.
3. **C3.** It incorporates both European and international regulatory frameworks, including UNESCO's Guidance and the EU's AI Act.
4. **C4.** It advances the notion of synthetic pedagogy, focusing on human instructional design and governance.

## 3. Materials and Methods

Scientists in recent years have extensively studied deepfakes and AI in education, but a significant gap remains in existing literature in terms of the subject. Analytically, most of the recent studies, such as those by [5,13], focus on broader and general applications of synthetic media in higher education or, in contrast, focus mainly on security-related issues, [58-60], without taking into account the multilingual MOOC contexts. Moreover, AI-generated videos and avatars, including [14], typically examine general properties and courses such as management and engagement via questions regarding multilingualism, cultural adaptation, or linguistic justice in general. Systematic scoping reviews of LLMs in education, such as [53, 60], highlight the ethical challenges of generative AI but do not examine synthetic media—videos, avatars, or cloned voices, specifically within MOOCs. As for the ethical approach of AI in education, following [52,61], we suggest that the generic architectural principles of AI are not actively applied when designing a multilingual MOOC that embeds synthetic media. In order to address this, one should try to incorporate the use of emerging technologies and technical tools such as avatars, voice cloning, and AI-generated videos with respect to the existing regulatory frameworks of UNESCO and the EU's AI Act.

As such, our article aims to develop a broader scoping review of multilingual MOOCs, focusing on both institutional reports and ethical and political articles. To achieve that, i.e., providing an interdisciplinary review, we considered [62]'s framework that highlights 6 key stages of action:

1. formulating research questions,
2. identifying relevant studies,
3. selecting studies,
4. mapping the data, and
5. summarising and reporting the results.

As such, using this framework, while taking into account its negatives, [63], we try to outweigh the cons by clearly defining our aims and objectives, i.e., our research questions, and showcase the importance of existing policy frameworks and regulations alike.

Moreover, the methodology of this review is aligned with the Joanna Briggs Institute (JBI) principles, as stated in the JBI Manual for Evidence Synthesis [64]. Additionally, as for the practical suggestions of [65]'s educational guide, we have implemented them by suggesting a simple but analytical step-by-step approach that remains consistent with JBI and PRISMA-ScR standards. Lastly, as for the review phase, we followed PRISMA Extension for Scoping Reviews (PRISMA-ScR) by [66], i.e., we followed the checklist and the respective notes for each part of the process.

In summary, our methodology was inspired and defined by [62,63], where we tried to adhere to the JBI framework and protocol recommendations so as to have a correct PCC (Population – Concept – Context) formulation and data charting, in regard to the transparent reporting requirements of PRISMA-ScR.

The methodological foundation combines the structural process defined by [62], the conceptual refinements proposed by [63], the specific procedural guidance of the JBI framework, particularly concerning protocol development, PCC formulation, and data charting, and the transparent reporting requirements of PRISMA-ScR. The review is conducted based on a predefined protocol, in accordance with JBI and PRISMA-ScR guidelines, defining the scope, purpose, research questions, databases, search strategy, inclusion and exclusion criteria, and data-mapping plan. The formulation of research questions follows JBI's PCC (Population–Concept–Context) structure: Population refers to students, teachers, MOOC designers, and higher education institutions; Concept refers to deepfake and synthetic-media technologies such as AI tutors, avatars, AI-generated videos, voice cloning, and speech-to-speech or speech-to-text translation for educational purposes; and Context refers to multilingual or international MOOCs and other forms of large-scale online learning, including open online university courses.

### 3.1 Information sources and search strategy

The search strategy was developed to capture both pedagogical and technological literature, to identify studies that address AI applications, such as deepfakes and synthetic media, within educational contexts and relate specifically to MOOCs, distance learning, and multilingualism. Searches were performed across a range of databases to ensure comprehensive coverage. IEEE Xplore was consulted for technological and computational applications, including speech-to-speech systems and avatar technologies. The ACM Digital Library was also extensively used to locate research in computational education, Human-Computer Interaction (HCI), and learning analytics. Scopus and the Web of Science Core Collection provided multidisciplinary and high-impact international peer-reviewed studies. Additional manuscript searches were carried out in SpringerLink and Taylor & Francis Online, which include extensive literature on educational technology, AI ethics, and the social sciences. Relevant IEEE Transactions and conference proceedings were also consulted for work on multimodal translation, while Google Scholar, ResearchGate, and Academia.edu served as a supplementary sources for reliable grey literature, including reports from UNESCO and official EU documentation such as the AI Act.

The search was limited to studies published between January 2020 and June 2025, a period that captures developments following the emergence of generative TN systems (GPT-3 and subsequent models) and the rapid expansion of research on LLMs in education. Both English and Greek language studies were eligible. In practice, almost all primary empirical and theoretical studies identified were in English, with Greek sources appearing only in a limited number of non-central references.

Search terms were defined inductively through a pilot search and preliminary review of titles and abstracts, drawing also on keywords from previous reviews on AI in education and deepfakes. An example of the search strategy used in Scopus is as follows:

```
query = (
  (
    "deepfake*" in content
    or "synthetic media" in content
    or "AI-generated video*" in content
    or "AI tutor*" in content
    or "avatar instructor*" in content
    or "voice cloning" in content
    or "speech-to-speech translation" in content
  )
  and
  (
    "MOOC*" in content
    or "massive open online course*" in content
    or "online course*" in content
    or "distance learning" in content
  )
  and
  (
    "multilingual" in content
    or "cross-linguistic" in content
    or "multilanguage" in content
    or "translation" in content
  )
  and
  (
    "education" in content
    or "higher education" in content
    or "e-learning" in content
  )
)
```

It is noted that in the final clause, the terms "education" in content and "higher education" in content were combined with an OR operator to maximise recall. Some studies describe their setting using the general term "education," whereas others specify "higher education" without repeating the broader term in searchable fields. Including both formulations ensured that we captured work situated explicitly in higher education as well as studies that addressed educational contexts more generally but still met our PCC criteria.

In order to keep the search process consistent across databases and to document the logic of the queries, we implemented a simple Java-based search model, which is described in the Appendix (Table 7, Figure 1). The program defines basic value objects for the time window, allowed languages, and concept blocks (e.g., synthetic media, MOOCs and distance learning, multilingual translation, educational context) and then assembles these into database-specific Boolean queries.

Rather than automating full access to proprietary APIs, the Java routines were used to generate and log the final query strings for each database, to standardise the combination of keywords across platforms, and to export the resulting records into structured tables (CSV files) with common fields (title, authors, year, source, DOI). These tables were then merged and de-duplicated based on DOI or title before manual screening. The purpose of this programming step was not to replace human judgment, but to make the search protocol transparent, reproducible, and technically documented beyond the narrative description.

Similarly, a generic Java programming language description of the above search can be found in the Appendix (Table 7, Figure 1). Lastly, it is noted that in more technologically oriented databases such as IEEE Xplore and ACM Digital Library, the strategy was adapted to include terms like *neural dubbing*, *multimodal translation*, and *speech-to-speech machine translation*, combined with terms relating to education, learning, or MOOCs.

The inclusion and exclusion criteria were aligned with the JBI PCC framework and with scoping review practices in artificial intelligence in education. These criteria are presented in Table 1.

*Table 1 Inclusion and Exclusion Criteria for the Comprehensive Review*

| Subject | Inclusion Criteria | Exclusion Criteria |
|---|---|---|
| Population | Students, teachers, MOOC designers, university institutions, and studies for learners in online higher education or lifelong learning | Primary/secondary students without connection to MOOCs; general population without educational context |
| Concept | Use or analysis of deepfake technologies, synthetic media, AI-generated video, avatars, voice cloning, speech-to-speech translation for educational purposes | Studies that examine exclusively malicious deepfakes (e.g., political propaganda, pornography) without an educational dimension |
| Context | MOOCs, open online courses, wide-ranging distance learning programs; multilingual or international environments | Purely in-person teaching with no online component; corporate training with no published academic documentation |
| Type of Study | Empirical (quantitative, qualitative, mixed), theoretical papers, scoping/systematic reviews, and institutional reports with explicit research/analytical methodology | Letters, opinions without methodology, journalistic articles, blog posts |
| Date | 2020–2025 | Before 2020 (unless they have a strong methodological influence - e.g., [62,63], which are only included in the methodological literature) |
| Language | English or Greek | Other languages without an available translation or English summary |
| Availability | Full text available (open access or via institutional access) | Abstract only, poster without full text |

## 3.2 Study selection and screening

The study selection process followed a series of sequential steps consistent with PRISMA-ScR and JBI guidelines. First, all results retrieved from the various databases were merged, and duplicates were removed. This was followed by a pre-screening of titles and abstracts using the predefined inclusion and exclusion criteria. Studies that met these criteria were then examined at the full-text level. Lastly, we manually checked the relevance in connection with our initial research questions to align with our initial aims and charted a generic characteristic for each study.

As such, during our research, either using a programming language and inclusion/exclusion criteria or manually interacting with each document, our study focused on the following:

- Total records identified across all databases: 514
- Records remaining after duplicate removal: 412
- Titles and abstracts screened: 412
- Full texts examined in depth: 94
- Final studies included in the mapping: 37

As such, Figure 2 provides an overview of the study selection process and Figure 3 summarises the screening and exclusion decisions in PRISMA-ScR format.

Lastly, based on the above, no standalone dataset was generated or stored during the review. All screening, verification, and inclusion decisions were conducted manually by the authors after executing Java-based search routines. Only temporary notes and downloaded PDFs were used for internal cross-checking and were not retained as a research dataset.

## 3.3 Data extraction and mapping

Following the JBI and PRISMA-ScR guidelines, initially, we focused on extracting data from our data to define some generic characteristics for each research subject. As such, during the review of each article, we focused on manually checking the following information:

- Basic bibliographic details, including authors, year, title, journal or conference venue, and DOI.

- Country or region of focus, or indication of an international context.

- Study type, classified as empirical quantitative, qualitative, mixed methods, theoretical, review, or policy report.

- Educational context, such as MOOC, SPOC, open course, or reference to a particular platform.

- Type of synthetic-media technology used (e.g., deepfake-style avatar tutors, AI-generated videos, voice cloning, speech-to-speech translation tools such as SEAMLESSM4T).

- Aim of the intervention or analysis, whether improving understanding, supporting language access, reducing costs, or critically examining the technology.

- Generic characteristics clearly or indirectly presented, such as educational level, demographics, and sample size (number of participants)

- Pedagogical properties and related values of key performance indexes, ranging from learning performance metrics to levels of engagement and social presence, or remarks about perceptions of authenticity.

- Ethical considerations, such as cultural bias and remarks from misinformation, and privacy concerns, are raised by students and teachers/tutors alike.

- Policy and regulatory framework suggestions spanning from Policy or governance recommendations and institutional guidelines to proposals for supporting AI literacy.

An overview of the data-charting process is presented in Table 2 below.

*Table 2 Overview of* the Data extraction and mapping rationale

| Category | Description |
|---|---|
| Identifiers | Author, year, journal, DOI |
| Context | Program type (MOOC, etc.), country/region, subject area |
| Technology | Specific type of synthetic media, technical approach, degree of automation |
| Pedagogical dimension | Educational objectives, theoretical framework, teaching methodology |
| Learning outcomes | Performance, engagement, satisfaction, and social presence indicators |
| Ethical/political dimension | Issues of privacy, transparency, governance, and inequalities |
| Recommended measures | Policy recommendations, guidelines, and implementation frameworks |

The table above focuses on presenting practical examples of AI tutors in higher education, [13], while also recommending real-life education LLMs, similar to [53].

The data charting and thematic synthesis followed a primarily inductive coding approach, in line with established scoping-review guidance. Initial codes were generated through close reading of the included studies, with attention to recurring concepts related to synthetic media, AI-enabled teaching practices, multilingual learning contexts, and governance or ethical considerations. These codes were iteratively refined and consolidated into higher-level categories, which informed the final data-charting structure presented in Table 2.

No formal inter-coder reliability statistics were calculated; however, the coding framework was reviewed and discussed iteratively within the author team to ensure conceptual coherence and internal consistency across categories. Coding was conducted manually using structured data-charting tables, rather than specialised qualitative analysis software, given the moderate size of the final study set and the descriptive–analytical aim of the scoping review. This approach allowed themes to be refined flexibly while maintaining transparency and traceability between the included studies and the resulting thematic categories.

Lastly, it is worth noting that during an initial pilot phase, a subset of studies was jointly reviewed by the author team in order to align interpretations of the data-charting categories and resolve coding ambiguities. The resulting refinements were incorporated into the shared coding tables before the full set of 37 studies was coded, so that the final themes reflected a calibrated understanding across researchers rather than the perspective of a single coder.

*3.4 Data analysis Rationale & Limitations*

As for the data analysis of our article, we focused on two main principles. Firstly, we focused on mapping the field in general, i.e., studying various recent papers from 2020-2025 regarding their type (theoretical, empirical case studies, generic reviews) that mentioned synthetic media technology in the area of education and case-specific studies in MOOCs, and open courses implementation. This quantitative overview enables comparisons with previous systematic and scoping reviews on synthetic media in higher education, including those by [53,67].

The second direction involved a thematic synthesis. Findings were inductively coded into categories aligned with the research questions, including technological typologies of synthetic media, pedagogical effects and learning experiences, ethical and cultural considerations, and policy or governance recommendations. These themes were subsequently grouped into higher-order categories in order to develop a conceptual model of *synthetic pedagogy*, organised along Technology–Pedagogy–Ethics–Policy axes. This dual approach of descriptive mapping and thematic analysis is consistent with related scoping reviews in educational AI, such as the systematic scoping review by [13,53,68] on deepfake-style AI tutors.

In line with recent JBI recommendations encouraging a shift from "consultation" to co-creation with knowledge users in scoping reviews, particular attention was given to incorporating perspectives from MOOC instructors, educational technology designers, and students with experience in online learning. These insights, gathered through informal discussions and pilot readings of the analytical framework, were not formally included in the dataset but contributed to refining and sensitising key concepts, especially regarding ethical and pedagogical issues.

Several methodological limitations must be acknowledged. Firstly, as is standard in scoping reviews and consistent with JBI's position, the review does not include a full critical appraisal of all empirical studies, since the goal is to map the field rather than assess effectiveness. Secondly, it is noted that focusing on recent, 2020 to 2205 literature might omit past important articles in the field; however, even though many older articles have technical merit, such as the ones by [62,63,67], we have considered them upon our manual screening of the newer ones. Thirdly, we have selected to search using the English language, which underrepresents the non-English references, but even though many interesting articles occurred especially for France, Portugal, and Nordic countries, we wanted this article to have a clear international orientation and results.

As such, even though some evident limitations can be argued, our approach and methodology try to promote international standards for generating a transparent and reproducible review in the subject of AI and synthetic media.

# 4. Results

*4.1 General Overview of Literature*

Our review focused on approximately 30 key references published from 2020 to 2025. Our overall screening process showcased that after 2022, there is a clear trend in the use of emerging technology, especially for text-to-image and video generative models (e.g., LLMs, avatars). Analytically, in terms of subject focus, approximately one-third of the studies analyse AI-generated or avatar-based instructional videos in higher education, often comparing learning outcomes and learner perceptions between human and AI-produced videos. A second group examines AI instructors or artificial pedagogical agents, with particular attention to their perceived social presence in online courses. A third group includes critical, review, and policy-oriented papers that address the ethical, regulatory, and pedagogical challenges associated with synthetic media and deepfakes in higher education.

Geographically, the literature is heavily concentrated in Europe, North America, and East Asia, with very limited evidence from other regions. This imbalance restricts our ability to understand global inequalities in access to multilingual MOOCs. A notable finding is that only a small number of studies explicitly investigate massive open online courses or multilingual learner populations. Most of the evidence derives from general online courses or from small-scale experimental settings. As a result, conclusions relevant to MOOCs largely reflect the transfer of insights from these adjacent educational environments.

The analysis of the studies confirms four main types of synthetic media use in higher education and MOOCs. The first category involves AI-generated instructional videos featuring avatars or deepfake-style tutors. The second concerns multilingual voice dubbing and voice-cloning applications designed to support linguistic accessibility in MOOCs. The third relates to conversational AI tutors based on LLMs, which provide personalised guidance or explanations. The fourth includes the use of synthetic media for generating examples, simulations, and activities aimed at developing critical thinking about deepfakes.

As such, to provide a clearer view of the empirical basis of this scoping review, Table 8 summarises the 37 studies included in the final mapping by educational context. Specifically, only a minority of these studies explicitly focus on large-scale MOOCs or multilingual MOOCs, while most empirical evidence derives from smaller online or blended higher-education courses, alongside a set of conceptual and policy-oriented contributions. As a result, this distribution supports a cautious transfer of findings from general online higher education to truly massive and multilingual MOOC settings. From Table 8, it is noted that, of our 37 studies, 11 were MOOC-based or MOOC-like, including general MOOCs, language/multilingual MOOCs, and work on MOOC big-data ethics. Similarly, 11 studies were based on smaller online or blended higher-education courses or pilots, mainly involving AI-generated videos, avatars, or LLM tutors. Lastly, 15 studies focused primarily on policy-oriented and conceptual findings, providing theoretical and scoping reviews and regulatory guidance regarding generative AI, synthetic media, and ethics in education for open and distance learning.

### 4.1.1 AI-generated and avatar-based educational videos

The study by [14] compared human- and AI-generated instructional videos in a management course. In an experiment involving 447 participants, the researchers used generative tools, including an LLM for script production, generative graphics, and avatar-based video creation, and found that learning outcomes, measured through test performance, were equivalent across conditions. However, participants reported a slightly more favourable subjective learning experience when viewing human-instructor videos.

Similarly, in [69], the authors present an online university course with 133 students that showed the presence of a humanoid artificial pedagogical agent embedded throughout course slides was positively associated with quiz performance and overall course satisfaction. The addition of avatars for student self-representation or peer representation did not further enhance performance, but it did influence students' lived learning experience and motivation.

In [13], the authors, in their systematic and mixed-methods review, document the emergence of "deepfake-style AI tutors" in fully online programmes, in which a synthetic avatar represents the instructor across multiple languages using lip-sync and voice-cloning technologies. The authors note that such tutors enable rapid production of multilingual content but also introduce new risks relating to impersonation and misrepresentation.

### 4.1.2 Multilingual translation and voice cloning & Professor conversation with LLM agents

At the technological level, advanced multimodal speech-to-text and speech-to-speech translation systems, such as Meta's models for seamless multilingual translation and speech synthesis, [70], enable the retention of an instructor's vocal identity, including voice colour and prosody, while changing the language of the course. This makes it possible for a MOOC to be delivered in multiple languages without re-recording, substantially reducing production costs. These systems have already been piloted in international online courses, primarily for subtitles, automatic translation, and audio description. The literature reports positive effects on comprehension for non-English-speaking students, although questions remain regarding cultural nuances, accent representation, and potential biases embedded in model training data.

The systematic review by [13,53,71] on the practical and ethical challenges of LLMs in education highlights the rapid expansion of chatbot-style tutors used to provide personalised learner support, including multilingual assistance. The main issue with most of these studies is that they are rather small in scale and are usually implemented in a laboratory classroom environment; thus, they limit their direct applicability to real case studies and scenarios of MOOCs.

### 4.1.3 Geographical distribution and country trends

The studies included in this review are not evenly distributed across regions. Most empirical research on AI-generated videos, avatar tutors, and AI-based translation tools comes from higher-income countries, mainly Europe, North America, and East Asia, with some additional work from other internationally oriented universities. Research that focuses on multilingual or language-related MOOCs appears more often in European contexts and in English-medium institutions that serve international or minority-language learners. By contrast, studies from regions such as Africa, Latin America, and parts of South and Southeast Asia are limited, even though many of these regions have rapidly growing student populations and heterogeneous linguistic landscapes. Policy and conceptual papers similarly reflect perspectives from European and North American institutions and international organizations, reinforcing a clear Global North focus in how synthetic media in education is theorized and governed.

These geographical patterns likely reflect a combination of factors, including the concentration of large MOOC platforms and research-intensive universities in higher-income countries, differential access to funding and infrastructure for educational AI projects, and the dominance of English-language publishing. Because of this imbalance, the findings of the present review mainly apply to well-resourced higher-education systems and MOOC ecosystems, where technological experimentation with synthetic media is already underway. Future research should therefore prioritize empirical studies on synthetic media and multilingual MOOCs in underrepresented regions and ideally combine bibliometric mapping with data on national MOOC participation and language policies, to understand whether current gaps are driven by research priorities, infrastructural constraints, or deeper inequalities in access to online higher education.

### 4.2 Educational Implications

The educational implications of this section are organized around the following central pedagogical dimensions: (a) learning outcomes, (b) engagement and social presence, (c) access and inclusion, (d) ethical and regulatory, (e) audiovisual protection, (f) regulatory frameworks, (g) findings, and a novice framework proposal.

#### 4.2.1 Learning Outcomes

In regard to the learning outcomes, they can be described as optimistic. Specifically, recent studies such as [14], showcase that no statistically significant difference was observed by students between human-generated and AI-generated instructional videos, even though, in terms of enjoyment and not educational process, participants preferred human videos over the AI ones. Similarly, in [69], the study suggested that the presence of an AI agent was considered as an asset regarding quiz performance, as students reported that they visually interacted with this pedagogical AI agent that assisted them in maintaining focus and interest on the taught material.

As such, the overall findings of our analysis suggest that synthetic or AI-generated videos do not contradict core learning outcomes and educational designs. Specifically, the advantages of human instructors tend to emerge more strongly in qualitative dimensions such as engagement, perceived connection, and sense of authenticity rather than in measurable performance indicators. With regard to multilingual MOOCs, the limited available evidence indicates that providing content in students' native languages through AI translation or voice cloning can enhance comprehension and confidence among learners with lower English proficiency, especially in technical subjects. However, there remains a notable lack of large-scale comparative studies that explicitly relate different multilingual implementation approaches to learning outcomes.

#### 4.2.2 Engagement, social presence, and credibility

Beyond performance, several studies concentrate on students' sense of social presence and the perceived credibility of the AI instructor. The authors in [72], found that students consider an AI instructor more credible when the voice used is human rather than robotic, and that social presence mediates the relationship between voice characteristics and perceived credibility. Credibility, in turn, was shown to predict students' intention to re-enroll in a course led by an AI instructor. Similarly to [69]'s findings, similarly indicate that attention directed toward the artificial agent and the instructor's perceived "cognitive presence" are positively associated with course satisfaction, although the emotional connection to the agent remains lower than would typically be expected with a human instructor.

At the level of scoping reviews, [13,73,74] conclude that deepfake-style tutors can enhance personalised guidance and increase the availability of feedback, but they continue to fall short in reproducing the full range of social and emotional cues that underlie trust and a sense of belonging. Overall, the interpretation of the findings converges on a central point:

the pedagogical value of synthetic media depends less on their technological sophistication and more on the way they are embedded within a coherent instructional design.

### 4.2.3 Access, inclusion, and multilingualism

In [55], OECD reports highlight that generative AI systems can serve as a mechanism for promoting inclusion, particularly for students with linguistic barriers or disabilities, by providing automatic translation, subtitles, and access to personalised learning material. The findings of this scoping review support this view: in several cases, students with low proficiency in the language of instruction report reduced anxiety when they can access avatar-based explanations in their native language, and accessibility features such as automatic subtitling and speech-to-text transcription are associated with improved participation in asynchronous courses.

At the same time, there is increasing concern about the emergence of a new form of "digital colonialism," whereby dominant English-language models generate translations and avatars that fail to capture the cultural and linguistic specificities of smaller or less-represented languages.

### 4.2.4 Ethical and regulatory dimensions

The second major focus of the scoping review concerns the ethical, social, and regulatory implications of synthetic media. Across studies and policy documents, four recurring themes emerge: transparency and informed consent; data protection and issues of voice and visual identity; academic integrity and risks of manipulation; and institutional accountability and AI governance.

Analytically, in [55], the study addresses transparency and informed consent by suggesting that students should, in all cases, be explicitly informed when instructional consent (i.e., the teacher) is not human but of a synthetic nature. Also, for this synthetic AI teacher, any use of personal data, including voice or image, requires documented consent. Similar conclusions are drawn in the analysis by [12,75] on deepfakes in higher education, which advocates for mandatory labelling of synthetic media in educational settings so as to safeguard institutional trust.

Within the studies examined in this scoping review, it was found that only a minority clearly disclosed to participants that the instructor was an avatar or that the material was synthetic. In most empirical studies, participants were informed only after completing the activity that the video had been AI-generated. This raises concerns about the extent to which such findings can be generalised to authentic educational contexts, where non-disclosure would likely be considered ethically problematic or even misleading.

### 4.2.5 Data, voice, and image protection

The use of deepfake and voice-cloning technologies for producing multilingual content requires the extensive collection of voice and visual data from teachers and, in some cases, from students. OECD policy documents and recent guidelines on ethical AI in education, e.g., [53] underline the importance of establishing clear ownership arrangements regarding who holds the rights to an instructor's synthetic avatar, limiting secondary uses such as commercial exploitation outside the educational context, and ensuring that collected data cannot be repurposed for non-consensual deepfake applications.

The study, [13], presents an interesting four-pillar governance framework that can be applied throughout cross-disciplinary domains:

- Transparency and Disclosure,
- Data Governance and Privacy,
- Integrity and Detection, and
- Ethical Oversight and Accountability,

All of the above are crucial as they can offer a solid foundation for the architectural principles of the MOOC institutional regulatory and policy-driven framework of using synthetic media.

### 4.2.6 Academic integrity, deception, and manipulation

One other issue that must be tackled is the academic integrity of generating deepfake avatars regarding potential violations, such as impersonations or oral examinations, or interviewing in general. This means that the development of fabricated videos can spike up metrics such as class attendance and mislead instructors and institutions into not putting work into producing actual thought and research questions in their everyday work. As such, studies like [53], highlight that LLMs can be used as a quick fix or a "shortcut" for completing assignments of any subject if their use is not monitored by strict, explicit rules and appropriate assessments designed with proper properties and criteria.

Moreover, multilingual MOOCs are evidently much more exposed to the above-mentioned risks. Specifically, by nature, their large and heterogeneous participant sample size poses a great challenge to detecting suspicious submissions and profiles. At the same time, synthetic avatars for accessibility purposes may blur the distinction between right and wrong or acceptable or not in real-life applications. As such, identifying misuses and legitimate uses is something complex that must be severely studied and examined for each case study and scenario.

### 4.2.7 Regulatory framework: EU AI Act, UNESCO, OECD

In terms of regulations and the existing regulatory reviews and frameworks for operation, the European Union's AI Regulation (AI Act) categorises education as a domain of high risk that requires swift action and measures in the areas of transparency, human oversight, and risk management. This happens as AI systems can severely influence educational progress and assessment alike. Similarly, OECD and Education International, [76], agree with this statement as it calls for action regarding specific "guidelines and safeguards" for achieving fair use and points to the teachers' and students' participation and overview of system designs in AI use in education as a crucial step for developing a balanced learning environment.

Taken together, these developments reveal a heterogeneous yet converging regulatory landscape. International organisations such as UNESCO, the OECD, and the Council of Europe, along with national authorities, increasingly highlight the importance of human-centred AI use, the protection of the rights of students and teachers, and the need to establish AI literacy as a fundamental competence for all stakeholders.

## 4.2.8 Comparative synthesis of findings and proposed policy framework

Based on the analysis of our most interesting sources, the scoping review synthesis, and our analysis, we present our findings in Table 3, which focuses on 3 categories of applications along with their associated pedagogical benefits and risks.

*Table 3 Categories of synthetic media use in multilingual MOOCs*

| Categories | Interesting Case Studies | Pedagogical Benefits | Ethical Risks |
|---|---|---|---|
| AI-generated instructional videos (avatars) | [14,69] | Equivalent learning outcomes, speed of material production, and adaptability | Reduced social presence, risk of "impersonal" learning, dependence on platforms |
| Deepfake-style AI tutors in multiple languages | [13] | Multilingual access, 24/7 support, and personalization | Impersonation, voice/image ownership, need for strict governance |
| LLM-based conversational tutors | [53,77] | Personalization, immediate feedback, and development of metacognitive skills | Academic integrity, dependence, opacity of models, and need for AI literacy |

Drawing on the table and the qualitative findings, the paper proposes a four-dimensional policy framework for the responsible use of deepfake and synthetic media in multilingual MOOCs.

1. **Pedagogy before technology**: Synthetic media should be embedded within a coherent instructional design—such as flipped learning or mastery learning—with a clear focus on learning outcomes rather than being used as a technological "gimmick."
2. **Mandatory transparency and clear labelling of synthetic content:** Platforms and courses should explicitly indicate which elements are synthetic (e.g., avatar, voice, or generated content), using accessible, readable, and multilingual disclosures.
3. **Institutional governance and risk assessment:** Institutions should adopt governance approaches aligned with frameworks such as those proposed by [13], and with OECD/UNESCO guidelines, [55,76]. This includes explicit procedures for data governance, detection mechanisms, incident response, and active participation of students and teachers in decision-making.
4. **AI literacy and understanding synthetic media use and applications:** Activities by educational program designs and institutions alike should be embedded to assist students to analyse, categorise, and evaluate synthetic media and promote active and informed participation.

# 5. Discussion

Based on the analysis of the previous section, synthetic media and deepfake use can be used for pedagogical purposes if trust and ethical considerations are taken into account. Analytically, recent studies such as [14] showcased that students are capable of learning effectively regardless of human or AI-generated video instructors. Similarly, [78], used an AI tutor and supported that student performance not only matches actual human teaching, but in terms of student performance, it is capable of even surpassing an active learning classroom environment.

Also, it is worth mentioning that the evaluation of educational tools, methods, and technology is not an easy task. Specifically, learning outcomes should not be linked with the evaluations of the technology, nor should the adoption of these AI systems shift the responsibility from the instructor to overall institutional and government policies and regulations. For example, recent studies, e.g., in intelligent tutoring systems (ITS) and AI-based learning tools, although they do not argue their effectiveness, are not yet properly studied, examined, and documented.

In our article, based on the above, we have the following observations in connection with the research questions defined at the beginning of this article:

- **RQ1 (forms of use):** We identify a continuum ranging from "light" applications, such as subtitles and dubbing, to more intensive interventions, including deepfake-style tutors that fully replace the visible presence of the instructor.
- **RQ2 (pedagogical implications):** Synthetic media can support learning, yet social presence and perceptions of authenticity remain fragile.
- **RQ3 (ethical and political dimensions):** As scale increases, particularly in MOOCs with multilingual audiences, concerns regarding transparency, privacy, and fairness become more pronounced.

In short, *technically, we may be ready; pedagogically and politically, not always*.

### *5.1 Towards a framework of "synthetic pedagogy" for multilingual MOOCs*

To move beyond the general assertion that "AI is a tool—it depends on how you use it," we need a more systematic understanding of the kind of pedagogy that emerges when the "teacher" is an avatar, the voice is cloned, and the language is automatically translated. By combining the findings of this review with recent major syntheses on AI tools in higher education, such as those by [79,80], we propose a *synthetic pedagogy* framework structured around four axes.

1. **Learning axis:** What dimensions of learning are being enhanced? Performance, metacognitive skills, critical thinking, or language confidence?
2. **Relationship axis:** How is the student–teacher–institution relationship reshaped when the instructor is "synthetic"?
3. **Justice axis:** Whom does technology include, and whom does it marginalise or overlook?
4. **Governance axis:** Who determines where, how, and under what safeguards synthetic media will be deployed?

To illustrate these considerations more concretely, Table 4 summarises three different "styles" of integrating synthetic media into multilingual MOOCs.

*Table 4 Overview of different styles for integration of synthetic media into multilingual MOOCs*

| Integration | Main Characteristics | Pedagogical Profile | Main Risks |
|---|---|---|---|
| Technological necessity | Last-minute dubbing/subtitling, avatars just to "look cool" | Mainly benefits in terms of accessibility; limited connection with activity planning | Low transparency, risk of errors, and cultural incompatibilities |
| Functional modernization | Systematic use of AI videos, avatars, and chat tutors with clear learning objectives | Improved consistency of material, possibility of personalization, and better student support | Risk of overdependence on platforms, shrinking role of teachers |
| Synthetic pedagogy criticism | Co-design with students, explicit teaching around synthetic media, and ethical discussion | AI also becomes a subject of learning, strengthening AI literacy | Requires time, staff training, and supportive policies |

The last category is important as it focuses on "AI Natives" and the need to treat students as interpreters of AI systems rather than mere users. As such, [81], shows us that universities must shift their focus to developing new rules and policies for AI literacy and the use of generative tools and synthetic media alike. The main goal is to apply these insights to the existing context of multilingual MOOCs to address whether an AI avatar should be permitted or not in a course. As such, one of the main issues is how we can design a learning experience where students will be able to determine what synthetic media is, know how to use it, and, most notably, care and understand how to critically evaluate it.

### 5.2 Ethics and political dimensions

In regard to the political level, following UNESCO's guidelines [55], generative AI in education is more or less a relatively unregulated field. Specifically, as noted by [82], although UNESCO acknowledges the risks involved in generative AI, it suggests that the technological advancement alone, supported by better, more complex, and holistic AI tools, will be able to address any problems that may occur. Specifically, this is referred to as "techno-solutionism", i.e., the tendency to define technology advancement as a self-evident part of the final solution while missing. As such, our review shows that there is a risk evident in multilingual MOOCs if a full AI-translated MOOC or other generic solutions to inequality and inclusion may arise. One should consider if the actual content is culturally appropriate and, overall, whether students understand its usage and operations. For example, following [79], regarding AI adaptation policies, the study presents 40 worldwide universities that state this advancement should be addressed with caution and optimism. Specifically, they suggest that AI tools can support teaching and learning alike, but they create problems as well, related to data protection issues (both in the EU General Data Protection Regulation (GDPR) regulation and worldwide) and ethical issues regarding academic integrity.

At the same time, the study, [83], pinpoints that AI in education is not a mere technological infrastructure. Analytically, it is connected with overall corporate interests, student data ecosystems, platform economies, and even more generic subjects such as the broader politics of edtech. All of these directly impact the synthetic media in MOOCs, i.e., they require different stakeholders to collaborate and talk about issues apart from the deepfake-style tutors and course, i.e., regarding data governance, content ownership, and institutional dependency.

Moreover, the study [84], also puts education in the spotlight. Specifically, it states that teachers and tutors are not simple users but may directly be involved in workload, professional identity, and accountability. For example, new policies on MOOC synthetic media from management may be difficult to implement and even more difficult to implement correctly if meaningful participation from them is not taken into account.

All of the above can be summed up to a simple question: *can AI democratise education?* The promise of multilingual MOOCs powered by synthetic media is compelling, i.e., that any student anywhere could, in parallel, access the same course, in their own language, delivered by the same avatar-instructor, thus eliminating inequalities. Unfortunately, recent studies are not so optimistic. For example, [85], argues that this statement is not only false but even misleading as the mere tools to participate, i.e., the access to devices, available time, and, in the long run, language skills, remain unchanged. Similarly, [86], takes it a step further by suggesting a new term: "bullshit universities". In detail, this term describes institutions that focus solely on automating teaching processes and operation costs while at the same time not reducing their cost or exclusion percentage. In this sense, a MOOC that is nothing more than an AI land field of synthetic media will be nothing more than an effort to increase the overall prestige and brand name of an organization, and not a mechanism for change.

Lastly, if we focus even further and consult the contrarians such as [87,88], one may picture a less idealized AI perspective where critical literacy is non-existent, and effort must be made into contemporary institutional governance and participatory design processes, to reduce existing inequalities.

### 5.3 Comparison of politics

To clarify the current landscape, it is useful to contrast—albeit in a schematic way—three levels of policy frameworks that directly or indirectly shape the use of synthetic media in MOOCs, as summarised in Table 5.

*Table 5 Comparison of key policy frameworks*

| Institution | Key Priorities | Approach Style | Question: "What does synthetic media MOOC stand for?" |
|---|---|---|---|
| **UNESCO** [55,82] | Human rights, inclusion, security, and AI literacy | High-level guidelines, persuasive discourse | Encourages responsible use; provides general principles for transparency and data protection, but without specific technical criteria for avatars/deepfakes |
| **EU AI Act** [17,56] | Risk management, transparency, and accountability for high-risk systems | Binding regulatory framework with risk categorization | Education is considered high risk; MOOC providers using synthetic media should demonstrate compliance (risk management, human oversight) |

| Institution | Key Priorities | Approach Style | Question: "What does synthetic media MOOC stand for?" |
|---|---|---|---|
| **Institutional Policies [41,79]** | Academic integrity, teaching quality, equality | Mix of regulations (what is allowed/prohibited) and usage strategies | Framework for how a university will decide whether to allow deepfake tutors, in which courses, and under what conditions |
| **Critical interventions [82,85,86]** | Social justice, critique of techno-solutionism, market role | Theoretical/critical analysis, often countering the "optimistic" narrative | They remind us that even with good regulations, who benefits and who is left out is a political issue, not just a technical one |

The table above does not provide a direct answer to the question of "what should be done," but it underscores a crucial point: synthetic media in MOOCs lie at the intersection of international policy, national regulation, institutional strategy, and critical academic debate. Any pedagogical use of such technologies must take all of these levels into account.

### *5.4 Relation to Existing Scoping and Systematic Reviews*

As such, several recent reviews provide important background on AI in higher education; however, they do not address our specific focus. In particular, reviews that examine AI applications and intelligent tutoring systems in higher education map a broad landscape of tools and use cases, yet most scoping and systematic reviews treat synthetic media only as a small subcategory and do not examine MOOCs or multilingual learning in depth (e.g., [67, 68, 87]). Similarly, other reviews focus on large language models and AI chatbots in education, [53, 60], or on AI-driven tutoring systems and learning tools more generally, [80, 89], without providing a detailed analysis of deepfake-style videos, avatar tutors, or speech-to-speech systems in open online courses.

In addition, work on AI ethics, governance frameworks, and regulatory policies is of particular relevance, [55, 56, 73, 76, 83, 87], but it rarely engages with concrete multilingual MOOC projects or case studies that involve synthetic-media-specific risks, such as voice and image cloning of instructors. As a result, the existing literature on synthetic media tends to focus either on generic risks in higher education, [75], or on deepfake-style AI tutors in institutional courses, [13], rather than on multilingual MOOCs.

Building on this literature, the present review differs in two key respects. First, it brings together synthetic media technologies, AI tutoring systems, large language models, and AI governance within a single analytical framework, areas that are usually examined separately. Second, it places these technologies explicitly in multilingual MOOC and open-course contexts, enabling a more systematic examination of how technological design, pedagogical practice, and regulatory considerations intersect in large-scale and linguistically diverse learning environments.

### 5.5 Limitations

First, the field itself is still evolving. Many practical uses of synthetic media in MOOCs have not yet been documented in academic journals; they remain at the level of technical reports, commercial white papers, or internal evaluations. As a result, our understanding is necessarily partial and likely more "academic" than the reality of current market practices.

Second, as noted in other major reviews of AI in education (e.g., [80,89]), most studies are small-scale, involve short-term interventions, and lack long-term follow-up concerning students' academic trajectories or professional development.

A further limitation concerns language. The review was restricted to English-language publications (with a small number of Greek-language sources), which has several implications. Countries and regions where synthetic media may already be widely used but not reported in English are underrepresented. Consequently, the picture of multilingual MOOCs largely reflects practices in English-speaking or internationally oriented universities rather than in local or regional institutions. The study, [85] and other critical theorists caution that such "blind spots" can lead to overly optimistic generalisations regarding the "democratic" potential of AI.

Moreover, a practical limitation is that we were unable to examine internal data from major MOOC platforms (e.g., Coursera, edX, FutureLearn), where large-scale experiments with synthetic media and multilingual support are likely already underway. These datasets are not publicly accessible, yet they play a decisive role in shaping the actual landscape of online education.

Finally, as summarised in Appendix Table 8, only a limited subset of the 37 mapped studies focus on large-scale MOOCs or explicitly multilingual MOOCs. Unfortunately, based on our study, most empirical evidence comes from smaller online or blended courses that tackle the subject from a more conceptual or policy-oriented lens. As such, this imbalance restricts our findings to the extent to which our results can be generalised to massive, open, multilingual learning environments and can be used to reinforce the need for future MOOC-specific research.

### 5.6 Next Steps and Suggestions

Based on the above, we can draw up a list of the main findings of this paper in Table 6, matching them up with the research questions.

Table 6 Research questions, key findings, and main recommendations

| Research Question | Conclusions | Proposals |
|---|---|---|
| What forms of deepfakes/synthetic media appear in multilingual MOOCs? | From "light" solutions (translation, subtitles) to fully avatar-based teachers and conversational tutors, truly massive and multilingual implementations are still limited. | Institutional mapping of existing practices and encouragement of pilot projects with clear pedagogical design. |

| Research Question | Conclusions | Proposals |
|---|---|---|
| What are the pedagogical implications? | Learning does not seem to be undermined – it is often equivalent to traditional methods. Social presence and authenticity remain vulnerable. | Design of "synthetic pedagogy" that combines avatars/AI tutors with highly interactive and critical reflection activities. |
| What are the ethical/political dimensions? | Strong concerns about transparency, consent, voice/image ownership, and big-data practices in MOOCs. | Adoption of institutional codes of conduct, alignment with international guidelines, and meaningful participation of students/teachers in governance. |

A further limitation of this review is that we did not systematically include external indicators of MOOC uptake or platform use at the country level. As a result, we cannot directly assess how differences in national MOOC ecosystems, digital infrastructure, or language policies may have influenced the geographical distribution of the studies in our corpus. It is likely that these factors partly explain why some regions are strongly represented in the literature, while others appear only rarely. For example, countries with well-established MOOC platforms, strong digital infrastructure, and supportive language or education policies may generate more research output in this area. A focused follow-up study that combines bibliometric analysis with country-level data on MOOC participation and platform usage would therefore be a useful extension of this work. Such an approach could help clarify whether the observed gaps in the literature reflect research priorities, infrastructural limitations, or deeper inequalities in access to online higher education.

### *5.7 Policy and institutional governance proposals*

Voice cloning technologies are central to many synthetic media applications for multilingual MOOCs, as they enable the same instructor to "speak" in multiple languages while retaining their vocal identity. The study by [90], shows that sharing a "digital voice copy" introduces complex trust dynamics and concerns about secondary uses. At the same time, the work of [91] demonstrates how a more modest technical solution, a hybrid-pipeline voice-cloning system adapted for low-resource environments, can support inclusion, provided that consent mechanisms, watermarking, and abuse-detection techniques are incorporated.

As such, based on these results and insights, many institutional policies must be developed. Firstly, any content using voice cloning or deepfake technology in general should be accordingly tagged and labelled. Secondly, tutors should have explicit agreements of fair usage that define the potential access to their voice/image or any audiovisual property. Lastly, similar to GDPR policy rationale, there should be a right of revocation guaranteed and always available, meaning that any instructor may withdraw consent or halt their AI avatar or any cloned property in existing or future courses.

At the MOOC and big-data level, the study by [91], is particularly revealing: across much research based on MOOC datasets, student consent is either unclear or insufficiently documented, and ethical considerations are frequently articulated only after the data have been analysed. To avoid replicating this pattern in deepfake and synthetic media applications, an *ethical-by-design* approach is required. This includes incorporating explicit notifications and consent options within the MOOC platform, designing learning analytics and AI technologies with minimal data capture, and assigning research ethics boards a substantive role in projects that combine synthetic media, MOOCs, and data analysis.

The policy proposals discussed in this chapter do not begin from zero. They build on established frameworks such as UNESCO's guidelines on generative AI in education; the EU AI Regulation (AI Act), which designates education as a high-risk domain; and institutional declarations such as the Universidad Internacional de La Rioja (UNIR) Declaration for an Ethical Use of Artificial Intelligence in Higher Education, which outlines principles for integrating AI in ways aligned with a university's mission.

A particular feature of multilingual MOOCs with synthetic media is that decisions are not made solely at the institutional level. They often involve inter-institutional collaborations and international platforms. Such an example is in study, [41], where the authors describe a multilingual Open Education Resources (OER) MOOC developed through university partnerships, and the work of [92], on AI applications within European university alliances illustrates that such policies can be co-developed at the network level.

## 6. Conclusions

This scoping review mapped 37 studies on synthetic media in higher education and open online learning, with a particular focus on multilingual MOOCs. As shown in Table 8, only a small number of these studies are genuinely MOOC-based or MOOC-like, including work on language and multilingual MOOCs and MOOC data ethics, while most empirical evidence comes from smaller online or blended higher-education courses and from conceptual or policy-oriented analyses. As a result, current knowledge about synthetic media in MOOCs is largely inferred from adjacent educational contexts rather than from systematic, large-scale evaluations of multilingual MOOCs themselves.

Across the empirical literature, a consistent finding is that AI-generated instructional videos and avatar-based tutors can support learning outcomes comparable to those achieved with human-recorded materials, particularly in terms of test scores and performance indicators [14, 39, 40]. At the same time, several studies report lower levels of perceived social presence, emotional connection, or trust when learners interact with fully synthetic instructors rather than visible human teachers [69, 72, 75]. Early pilot studies on multimodal and speech-to-speech translation systems, including large multilingual models such as SeamlessM4T, indicate strong potential for improving access for non-English-speaking learners by enabling a single instructor to reach multiple language communities without repeated content production [34–36]. Together, these findings suggest that synthetic media are especially well-suited for scaling instructional content and supporting multilingual access, but that careful pedagogical design is required to maintain authenticity, trust, and a clear sense of human presence.

The review also shows that ethical, cultural, and regulatory issues are central to the educational use of synthetic media rather than secondary concerns. Research on AI ethics in education and generative AI highlights the importance of transparency, informed consent, explainability, and shared responsibility when synthetic media are used in learning environments [52, 53, 91]. International and regional policy frameworks, including UNESCO's Guidance for Generative AI in Education and Research and the EU Artificial Intelligence Act, provide relevant principles on human oversight, risk management, and disclosure that directly apply to deepfake-style videos, voice cloning, and multilingual translation in MOOCs and other online courses [55, 56]. Case studies on multilingual MOOCs and open educational resources further illustrate how issues of privacy, data use, and linguistic justice become concrete in practice, particularly for minority and non-dominant languages [33, 44, 92].

Overall, the findings of this review point towards the need for a balanced and cautious integration of synthetic media in multilingual MOOCs. Rather than replacing human instructors, avatar-based videos, cloned voices, and multilingual dubbing should be understood as tools for extending human-designed learning environments, reducing production costs, and improving access for linguistically diverse learners. At the same time, risks related to impersonation, deepfake deception, and intensified data surveillance require explicit institutional policies, clear disclosure of synthetic content, and participatory governance involving both educators and students. Future research should prioritise MOOC-specific and large-scale evaluations of synthetic media, examine their use in underrepresented regions and languages, and combine pedagogical experiments with analyses of national MOOC ecosystems and regulatory contexts. Only by connecting technological innovation with pedagogical theory, ethical reflection, and attention to multilingual realities can synthetic media contribute to more inclusive and transparent MOOCs rather than more automated and unequal ones.

# 7. Appendix

*Table 7 A generic Java programming language representation of our search criteria and UML objects*

```java
public class DeepfakeMoocSearchModel {
    /* ====== ENUMS (UML: simple value types) ====== */
    public enum Language { ENGLISH, GREEK }
    public enum ConceptBlockType {
        AI_SYNTHETIC_MEDIA, MOOC_DISTANCE_LEARNING,
        MULTILINGUAL_TRANSLATION, EDUCATION_CONTEXT
    }
    public enum DatabaseType {
        SCOPUS, WEB_OF_SCIENCE_CORE_COLLECTION, IEEE_XPLORE,
        ACM_DIGITAL_LIBRARY, SPRINGER_LINK, TAYLOR_FRANCIS,
        GOOGLE_SCHOLAR, RESEARCH_GATE, ACADEMIA_EDU
    }
    /* ====== BASIC VALUE OBJECTS ====== */
    public static class DateRange {
        public String startDate;   // e.g. "2020-01-01"
        public String endDate;     // e.g. "2025-06-30"
        public DateRange(String startDate, String endDate) {
            this.startDate = startDate; this.endDate = endDate;
        }
    }
    public static class KeywordBlock {
        public ConceptBlockType type;
        public List<String> terms;        // e.g. ["deepfake*", "synthetic media", ...]
        public String logicalOperator;    // usually "OR"
        public boolean required;          // true = must be present in query
        public KeywordBlock(ConceptBlockType type, List<String> terms,
                            String logicalOperator, boolean required) {
            this.type = type; this.terms = terms;
            this.logicalOperator = logicalOperator; this.required = required;
        }
    }
    public static class DatabaseSource {
        public DatabaseType type;
        public List<String> rolesKeywords; // e.g. ["multidisciplinary", "high_impact", "peer_reviewed"]
        public DatabaseSource(DatabaseType type, List<String> rolesKeywords) {
            this.type = type; this.rolesKeywords = rolesKeywords;
        }
    }
    /* ====== CORE SEARCH CRITERIA (UML: central class) ====== */
    public static class SearchCriteria {
        public DateRange temporalScope;
        public List<Language> allowedLanguages;
        public List<KeywordBlock> conceptBlocks;
        public SearchCriteria(DateRange temporalScope,
                              List<Language> allowedLanguages,
                              List<KeywordBlock> conceptBlocks) {
            this.temporalScope = temporalScope;
            this.allowedLanguages = allowedLanguages;
            this.conceptBlocks = conceptBlocks;
        }
    }
    /* ====== RATIONALE AS KEYWORD TAGS ====== */
    public static class SearchRationale {
        public List<String> temporalKeywords;    // e.g. ["post_GPT3", "LLM_education_wave"]
        public List<String> languageKeywords;    // e.g. ["primary_English", "secondary_Greek"]
```

```java
        public List<String> intersectionKeywords;  // e.g. ["AI_synthetic_media", "MOOCs", "multilingual"]
        public List<String> inclusionRules;        // e.g. ["all_blocks_AND", ...]
        public List<String> exclusionRules;        // e.g. ["exclude_purely_technical", ...]
        public List<String> policyKeywords;        // e.g. ["UNESCO_reports", "EU_AI_Act"]
        public SearchRationale() {
            this.temporalKeywords = new ArrayList<>();
            this.languageKeywords = new ArrayList<>();
            this.intersectionKeywords = new ArrayList<>();
            this.inclusionRules = new ArrayList<>();
            this.exclusionRules = new ArrayList<>();
            this.policyKeywords = new ArrayList<>();
        }
    }
    /* ====== PROTOCOL (UML: aggregates criteria, sources, rationale) ====== */
    public static class SearchProtocol {
        public SearchCriteria criteria;
        public List<DatabaseSource> sources;
        public SearchRationale rationale;
        public SearchProtocol(SearchCriteria criteria,
                              List<DatabaseSource> sources,
                              SearchRationale rationale) {
            this.criteria = criteria;
            this.sources = sources;
            this.rationale = rationale;
        }
        public String buildBooleanQuery(String fieldName) {
            StringBuilder query = new StringBuilder("(");
            for (int i = 0; i < criteria.conceptBlocks.size(); i++) {
                KeywordBlock block = criteria.conceptBlocks.get(i);
                StringBuilder blockBuilder = new StringBuilder("(");
                for (int j = 0; j < block.terms.size(); j++) {
                    String term = block.terms.get(j);
                    blockBuilder.append(fieldName).append("(\"").append(term).append("\")");
                    if (j < block.terms.size() - 1) {
                        blockBuilder.append(" ").append(block.logicalOperator).append(" ");
                    }
                }
                blockBuilder.append(")");
                query.append(blockBuilder);
                if (i < criteria.conceptBlocks.size() - 1) query.append(" AND ");  // all blocks mandatory
            }
            query.append(")");
            return query.toString();
        }
    }
    /* ====== FACTORY: BUILD OUR SPECIFIC PROTOCOL ====== */
    public static SearchProtocol buildDeepfakeMoocMultilingualProtocol() {
        DateRange range = new DateRange("2020-01-01", "2025-06-30");
        List<Language> langs = Arrays.asList(Language.ENGLISH, Language.GREEK);

        KeywordBlock aiBlock = new KeywordBlock(
                ConceptBlockType.AI_SYNTHETIC_MEDIA,
                Arrays.asList(
                        "deepfake*", "synthetic media", "AI-generated video*",
                        "AI tutor*", "avatar instructor*", "voice cloning",
                        "speech-to-speech translation"
                ),
                "OR", true
        );
        KeywordBlock moocBlock = new KeywordBlock(
                ConceptBlockType.MOOC_DISTANCE_LEARNING,
                Arrays.asList("MOOC*", "massive open online course*", "online course*", "distance learning"),
                "OR", true
        );

        KeywordBlock multilingualBlock = new KeywordBlock(
                ConceptBlockType.MULTILINGUAL_TRANSLATION,
                Arrays.asList("multilingual", "cross-linguistic", "multilanguage", "translation"),
                "OR", true
        );

        KeywordBlock educationBlock = new KeywordBlock(
                ConceptBlockType.EDUCATION_CONTEXT,
                Arrays.asList("education", "higher education", "e-learning"),
                "OR", true
        );

        SearchCriteria criteria = new SearchCriteria(
                range, langs, Arrays.asList(aiBlock, moocBlock, multilingualBlock, educationBlock)
        );
        List<DatabaseSource> dbSources = Arrays.asList(
                new DatabaseSource(DatabaseType.SCOPUS, Arrays.asList("multidisciplinary", "high_impact", "peer_reviewed")
                ),
                new DatabaseSource(DatabaseType.WEB_OF_SCIENCE_CORE_COLLECTION, Arrays.asList("multidisciplinary", "core_collection", "peer_reviewed")
                ),
                new DatabaseSource(DatabaseType.IEEE_XPLORE, Arrays.asList("technical", "synthetic_media", "speech_systems", "avatar_technologies")
                ),
                new DatabaseSource(DatabaseType.ACM_DIGITAL_LIBRARY, Arrays.asList("HCI", "learning_analytics", "computational_education")
                ),
```

```
            new DatabaseSource(DatabaseType.SPRINGER_LINK, Arrays.asList("educational_technology", "AI_ethics", "social_sciences")
            ),
            new DatabaseSource(DatabaseType.TAYLOR_FRANCIS, Arrays.asList("education", "social_policy", "AI_related")
            ),
            new DatabaseSource(DatabaseType.GOOGLE_SCHOLAR, Arrays.asList("grey_literature", "broad_discovery")
            ),
            new DatabaseSource(DatabaseType.RESEARCH_GATE, Arrays.asList("grey_literature", "author_versions")
            ),
            new DatabaseSource(DatabaseType.ACADEMIA_EDU, Arrays.asList("grey_literature", "author_versions")
            )
        );
        SearchRationale r = new SearchRationale();
        r.temporalKeywords.addAll(Arrays.asList("window_2020_2025", "post_GPT3", "LLM_education_focus"      ));
        r.languageKeywords.addAll(Arrays.asList(
            "include_English", "include_Greek",
            "primary_English_empirical", "secondary_Greek_contextual"                                      ));
        r.intersectionKeywords.addAll(Arrays.asList(
            "AI_synthetic_media", "MOOC_distance_learning", "multilingual_cross_linguistic"                ));
        r.inclusionRules.addAll(Arrays.asList(
            "all_concept_blocks_required", "logical_AND_between_blocks", "logical_OR_within_blocks" ));
        r.exclusionRules.addAll(Arrays.asList(
            "exclude_purely_technical_no_education", "exclude_purely_pedagogical_no_AI"                    ));
        r.policyKeywords.addAll(Arrays.asList(
            "UNESCO_guidance_generative_AI", "EU_AI_Act", "AI_ethics", "data_protection"                   ));
        return new SearchProtocol(criteria, dbSources, r);
    }
    /* ====== EXAMPLE MAIN (BUILD + BOOLEAN QUERY) ====== */
    public static void main(String[] args) {
        SearchProtocol protocol = buildDeepfakeMoocMultilingualProtocol();
        String scopusQuery = protocol.buildBooleanQuery("TITLE-ABS-KEY");
        System.out.println(scopusQuery);
}}
```

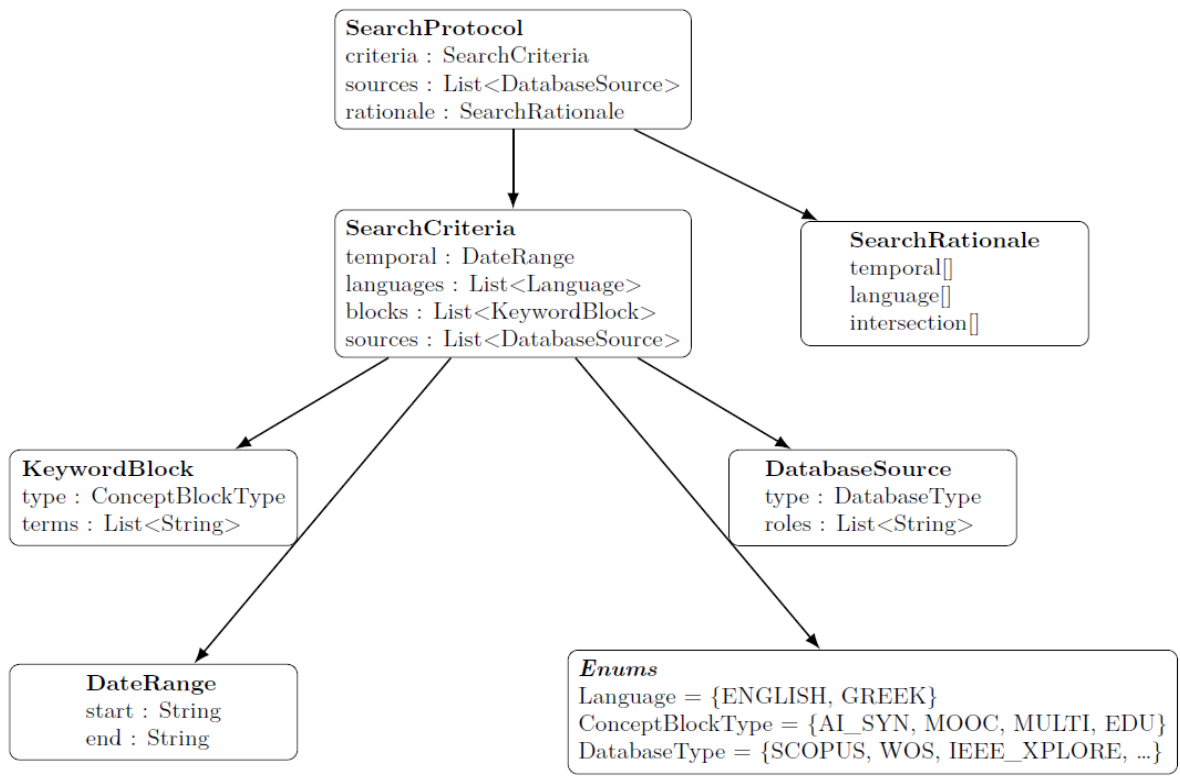

*Figure 1 UML diagram of Java code used for our scoping review search criteria*

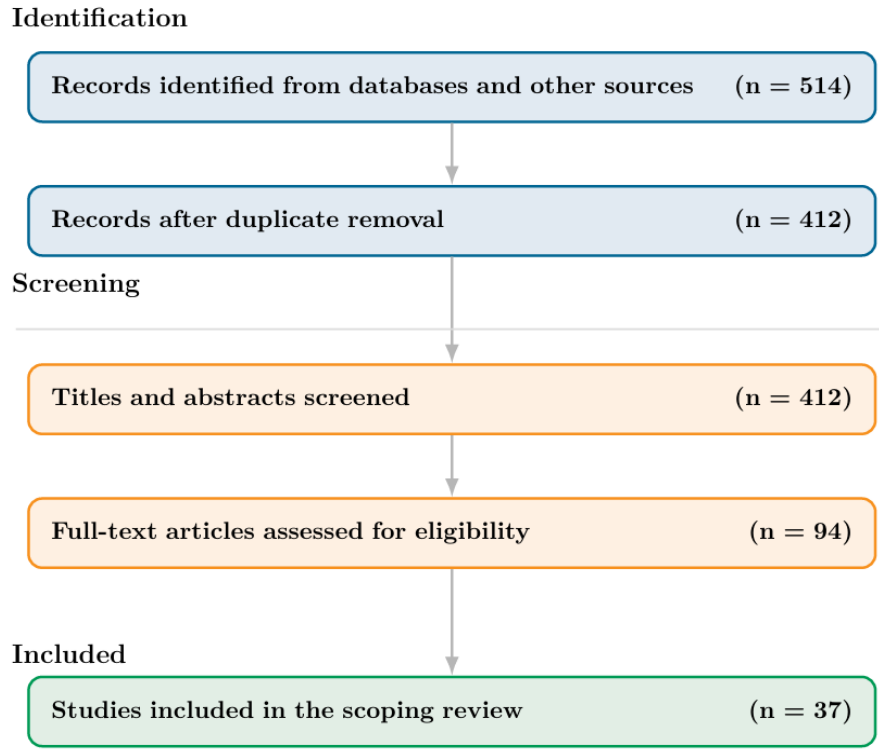

Figure 2 PRISMA-ScR flow diagram for study selection (overview)

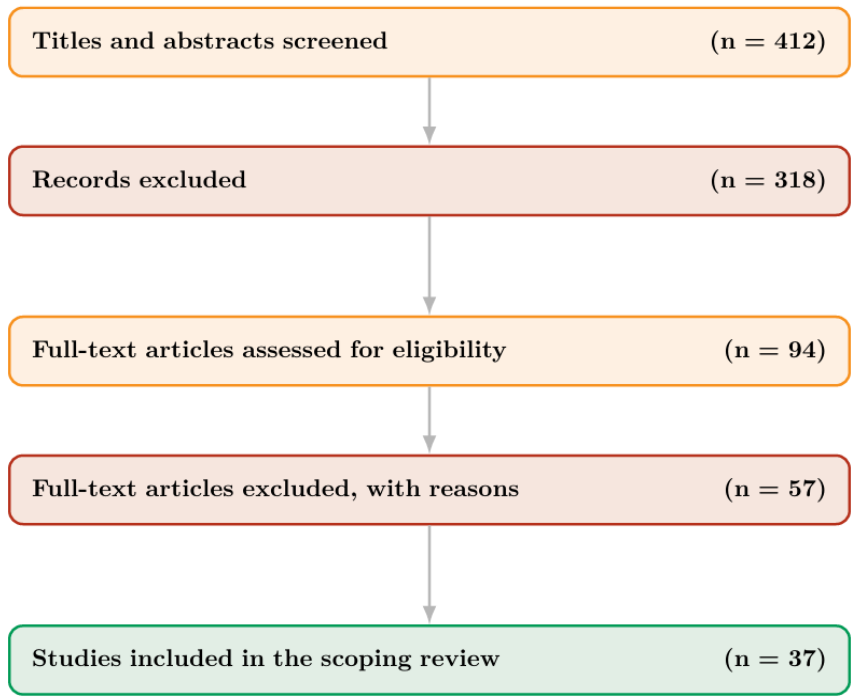

Figure 3 PRISMA-ScR flow diagram for study selection (screening and exclusions)

*Table 8 MOOC-based studies explicitly focus on massive open online courses; smaller online or blended courses include course-level or pilot empirical studies; policy or conceptual studies address frameworks, reviews, ethics, or governance. "Mixed" denotes studies combining conceptual analysis with limited empirical elements.*

| Reference Id | Multilingual or MOOC_focus | Technology or AI focus | Smaller Online or Blended Course | Policy Conceptual |
|---|---|---|---|---|
| [2] | MOOC | Deepfake-based video generation in MOOC content (Lessonable) | No | No |
| [3] | Language MOOCs (MOOC) | Language MOOCs in Mainland China, sustainable development | No | Yes |
| [4] | MOOC | Practices, trends, and challenges of MOOCs in HE | No | Yes |
| [6] | MOOC | Systematic review & bibliometric analysis of MOOCs in engineering education | No | Yes |
| [11] | MOOC | Big data–driven sentiment analysis of Chinese MOOC reviews | No | Partly |
| [31] | MOOC | Conceptual framework on hegemonic design bias in MOOCs | No | Yes |
| [32] | MOOC | Cultural inclusiveness of MOOC learning designs | No | Yes |
| [33] | Multilingual MOOC | Diversity, exclusion, and minority language representation in MOOCs | No | Yes |
| [45] | MOOC | Bridging MOOCs, smart teaching, and AI towards unified pedagogy | No | Yes |
| [91] | MOOC | Ethical research in MOOC big-data studies (Massive omission of consent) | No | Yes |
| [92] | Multilingual MOOC | Multilingual OER MOOC, production and usage in university cooperation | No | Yes |
| [5] | No (online HE context) | GenAI-supported assessment (AI Assessment Scale, AIAS) | Yes | Partly |
| [10] | No (HE courses) | Generative AI in education: usability, ethics, communication effectiveness | Yes (study context) | Partly |
| [14] | No (university course) | Comparing human-made vs AI-generated teaching videos (learning effects) | Yes | No |
| [39] | No (teacher education course) | AI-generated instructional videos in problem-based learning | Yes | No |
| [40] | No (HE course) | AI-generated short videos vs paper materials, effect of student acceptance | Yes | No |
| [41] | No (various pilots) | AI-generated courses and videos, pilots in AI-driven education | Yes | Partly |
| [42] | No (language learning course) | AI vs human translation, AI-generated bilingual captions for learning | Yes | No |
| [69] | No (online HE course) | Artificial pedagogical agent as teacher, visual learner avatars | Yes | No |
| [72] | No (online course) | AI instructor, social presence, and voice features | Yes | No |
| [74] | No (journalism course) | Chatbot style guide vs traditional stylebook for copy-editing | Yes | No |

| [78] | No (university course) | AI tutoring vs in-class active learning (LLM-based tutor) | Yes | No |
|---|---|---|---|---|
| [12] | No (HE) | Deepfakes and HE: scoping review and research agenda | No | Yes |
| [13] | No (HE) | Deepfake-style AI tutors: mixed-methods review & governance framework | No | Yes |
| [25] | No (education generally) | Generative AI for personalized learning: foundations and trends | No | Yes |
| [26] | No (open and distance learning) | Generative AI ethics in open and distance learning (ChatGPT) | No | Yes |
| [27] | No (HE) | Generative AI and education: digital pedagogies, teaching innovation, design | No | Yes |
| [44] | No (HE) | Linguistic justice and GenAI biases in an EMI university (case study) | Yes | Yes |
| [52] | No (education generally) | Ethics of AI in education: towards a community-wide framework | No | Yes |
| [53] | No (education generally) | Practical and ethical challenges of LLMs in education (systematic scoping review) | No | Yes |
| [55] | No (global education systems) | Guidance for generative AI in education and research (UNESCO) | No | Yes |
| [56] | No (AI regulation, including education) | EU Artificial Intelligence Act, legal analysis incl. education as high-risk | No | Yes |
| [57] | No (HE regulation) | Regulatory framework for AI use in European HE | No | Yes |
| [60] | No (education generally) | Umbrella review on ChatGPT and AI chatbots in education | No | Yes |
| [67] | No (HE) | Systematic review of AI applications in HE | No | Yes |
| [73] | No (education generally) | Ethical and regulatory challenges of generative AI in education (systematic review) | No | Yes |
| [76] | No (education systems) | Opportunities, guidelines, and guardrails for AI in education | No | Yes |

*\*HE stands for Higher Education*

Funding: The APC was funded by Academia AI and Applications journal editors as part of a special issue/call for papers.

Author contributions: A.G. and T.V.; methodology, A.G. and E.C.; software, N.N.; validation, A.G., E.C. and N.N.; formal analysis, A.G.; investigation, A.G. and T.V.; resources, T.V.; data curation, A.G.; writing—original draft preparation, A.G., E.C. and T.V.; writing—review and editing, T.V. and A.G.; supervision, A.G. and T.V.; project administration, A.G. and T.V.

All authors have read and agreed to the published version of the manuscript.

Conflict of interest: The author(s) declare no conflict of interest.

Data availability statement: Not applicable. No standalone dataset was generated or retained during this manuscript. All screening, verification, and inclusion decisions were conducted manually based on the search outputs, and temporary working files (notes/PDFs) were not retained as a research dataset.

Institutional review board statement: Not applicable.

Informed consent statement: Not applicable.

Sample availability: The author(s) declare that no physical samples were used in this study.

Supplementary materials: Not applicable.

Additional information